\documentclass[manuscript,usenatbib]{aastex}

\usepackage[usenames]{color}

\shorttitle{Has a star enough energy to excite the thousand
  of modes observed with CoRoT?}
\shortauthors{Moya and Rodr\'iguez-L\'opez}

\begin{document}

%% LaTeX will automatically break titles if they run longer than
%% one line. However, you may use \\ to force a line break if
%% you desire.

\title{Has a star enough energy to excite the thousand
  of modes observed with CoRoT?}

%% Use \author, \affil, and the \and command to format
%% author and affiliation information.
%% Note that \email has replaced the old \authoremail command
%% from AASTeX v4.0. You can use \email to mark an email address
%% anywhere in the paper, not just in the front matter.
%% As in the title, use \\ to force line breaks.

\author{A. Moya\altaffilmark{1}}
\affil{Instituto de Astrof\'isica de Andaluc\'ia, IAA - CSIC,
    Granada, Spain E-18008}
\email{amoya@cab.inta-csic.es}

\and
\author{C. Rodr\'iguez-L\'opez\altaffilmark{2}}

\affil{Laboratoire d'Astrophysique de Toulouse-Tarbes, CNRS,
     Universit\'e de Toulouse. Toulouse, France F-31400}

%% Notice that each of these authors has alternate affiliations, which
%% are identified by the \altaffilmark after each name.  Specify alternate
%% affiliation information with \altaffiltext, with one command per each
%% affiliation.

\altaffiltext{1}{Laboratorio de Astrof\'isica Estelar y Exoplanetas, LAEX-CAB (INTA-CSIC), PO BOX 78, 28691 Villanueva de la Ca\~nada, Madrid, Spain}
\altaffiltext{2}{Universidade de Vigo, Dpt. F\'isica Aplicada, Vigo, Spain E-36210}

%% Mark off your abstract in the ``abstract'' environment. In the manuscript
%% style, abstract will output a Received/Accepted line after the
%% title and affiliation information. No date will appear since the author
%% does not have this information. The dates will be filled in by the
%% editorial office after submission.

\begin{abstract}

The recent analyses of the light curves provided by CoRoT have
revealed pulsation spectra of unprecedented richness and precision, in
particular, thousands of pulsating modes, and a clear distribution of
amplitudes with frequency.  In the community, some scientists have
started doubting about the validity of the classical tools to analyze
these very accurate light curves.

This work provides the asteroseismic community with answers to this
question showing that (1) it is physically possible for a star
to excite at a time and with the observed amplitudes such a large
number of modes; and (2) that the kinetic energy accumulated in all
those modes does not destroy the equilibrium of the star.
Consequently, mathematical tools presently applied in the
analyses of light curves can a priori be trusted. This conclusion is
even more important now, when a large amount of space data coming from
Kepler are currently being analyzed.

The power spectrum of different stellar cases, and the non-adiabatic
code GraCo have been used to estimate the upper limit of the energy
per second required to excite all the observed modes, and their total
kinetic energy. A necessary previous step for this study is to infer
the relative radial pulsational amplitude from the observed
photometric amplitude, scaling our linear pulsational solutions to
absolute values. The derived upper limits for the required pulsational
energy were compared with 1) the luminosity of the star; and 2) the
gravitational energy. We obtained that both upper energy limits
are orders of magnitude smaller.

\end{abstract}

%% Keywords should appear after the \end{abstract} command. The uncommented
%% example has been keyed in ApJ style. See the instructions to authors
%% for the journal to which you are submitting your paper to determine
%% what keyword punctuation is appropriate.


%% From the front matter, we move on to the body of the paper.
%% In the first two sections, notice the use of the natbib \citep
%% and \citet commands to identify citations.  The citations are
%% tied to the reference list via symbolic KEYs. The KEY corresponds
%% to the KEY in the \bibitem in the reference list below. We have
%% chosen the first three characters of the first author's name plus
%% the last two numeral of the year of publication as our KEY for
%% each reference.

%% Authors who wish to have the most important objects in their paper
%% linked in the electronic edition to a data center may do so by tagging
%% their objects with \objectname{} or \object{}.  Each macro takes the
%% object name as its required argument. The optional, square-bracket
%% argument should be used in cases where the data center identification
%% differs from what is to be printed in the paper.  The text appearing
%% in curly braces is what will appear in print in the published paper.
%% If the object name is recognized by the data centers, it will be linked
%% in the electronic edition to the object data available at the data centers
%%
%% Note that for sources with brackets in their names, e.g. [WEG2004] 14h-090,
%% the brackets must be escaped with backslashes when used in the first
%% square-bracket argument, for instance, \object[\[WEG2004\] 14h-090]{90}).
%%  Otherwise, LaTeX will issue an error.

\section{Introduction}

The first data provided by CoRoT for $\delta$ Scuti stars made clear
that the larger the precision of the observations, the larger the
number of observed modes \citep{ennio,antonio}. In the case, for
example, of HD\,174936 \citep{antonio}, a $\delta$ Scuti target of the
initial CoRoT run, the number of frequencies extracted from the
observed light curves with the traditional tools can be of the order
of thousands. The same can be said from other CoRoT $\delta$ Scuti and
$\gamma$ Doradus targets. In such a scenario, the question if
all these frequencies are real or not should be a previous step to any
modeling attempt. To answer this, our aim is to check if a $\delta$
Scuti or a $\gamma$ Doradus star has enough energy to excite such a
large amount of modes simultaneously and with the observed amplitudes,
for which we need to know the real energy of the modes.

Up to our knowledge, the only existing studies about the real energy
of the modes, through characterisation of their amplitudes, are for
stochastic pulsators (as solar-like stars) which were done through a
description of the non-linear effects in the pulsation equations
\citep{reza03,kevin08}. In the case of thermodynamic pulsators, as
those driven by a $\kappa$ mechanism, this kind of studies have not
yet been attempted, since the non-adiabatic studies have been done in
the linear approximation \citep{osaki}. Therefore, the solutions
yielded by the pulsation codes are not absolute, as they are affected
by a constant indefinite factor, and only ratios of observed and
theoretical quantities can be compared.

We scaled the solutions given by the non-adiabatic linear pulsational
code GraCo \citep{moya04,moya08} to absolute values using the
observational amplitudes provided by space and ground-base data,
as described below. In this way, we were able to compare the derived
absolute values of the required pulsational energy to simultaneously
excite all the observed modes, with the real energy available for
exciting the modes.

In our study we use one of the models best reproducing the
ground-based and space observations obtained for the $\delta$ Scuti
HD\,174936 \citep[see][]{antonio}, and a model at the centre of the
observed HR photometric error box for the hybrid $\delta$ Scuti -
$\gamma$ Doradus star HD\,49434. In order to check the consistency of
our procedure, we use a representative model of the largest
amplitude pulsator known, Classical Cepheid stars presenting
variations of the order of magnitudes, and verified that it is
energetically stable under its pulsations.

%An example of a
%clasical Cepheid star pulsation with the maximum amplitude known has
%been used to test our procedure.

% In order to test the consistency of
%our procedure we have also verified if the largest amplitude pulsator
%known, a clasical Cepheid star presenting variations of the order of
%the mag, is energically stable.

\section{Scaling linear solutions to absolute values}

To determine the energy needed by the system to unstabilize all of the
observed modes, we have to face the problem that most of the
pulsational codes, and in particular the non-adiabatic ones, use the
linear approximation to solve the system of differential
equations. Aside from non-linear physical effects, negligible for
small pulsational amplitudes, this means that every eigenfunction of
an oscillation mode is known except for a multiplicative constant,
i.e., if $ f_i (x) $ are the solutions of the set of differential
equations for a given mode, $ Af_i (x) $, $A$ being a constant, are
also solutions. Therefore, before doing any study about the global
properties of the star, we have to determine the value of this
constant ($A_{real}$) scaling the linear solutions of each mode to the
real values.

In the linear approximation, the problem of this indefinite constant
is solved by scaling all the solutions of the pulsation equations to
an arbitrary normalization condition. Most pulsation codes use as
normalization condition

\begin{equation}
{\xi_r\over r}=1\;\;\; \mathrm{at} \;\;\; r=R
\label{norm}
\end{equation}

\noindent where $\xi_r/r$ is the radial eigenfunction
\citep{dziembowski71}. The real pulsation amplitude on the surface is
given by

\begin{equation}
{\delta R\over R}=A_{real}{\xi_r(r=R)\over R}
\end{equation}

\noindent And following Eq.~\ref{norm}

\begin{equation}
{\delta R\over R}=A_{real}
\end{equation}

\noindent Therefore, the multiplicative constant $A_{real}$, scaling
all the linear solutions to the real values of the pulsation, may be
obtained through the oscillation amplitude of the relative radius at
the surface of the star. This value can be deduced from the
oscillation amplitude of the observed luminosity.

\begin{figure*}
     \begin{tabular}{cc}
 \resizebox{0.48\linewidth}{6.5cm}{\includegraphics{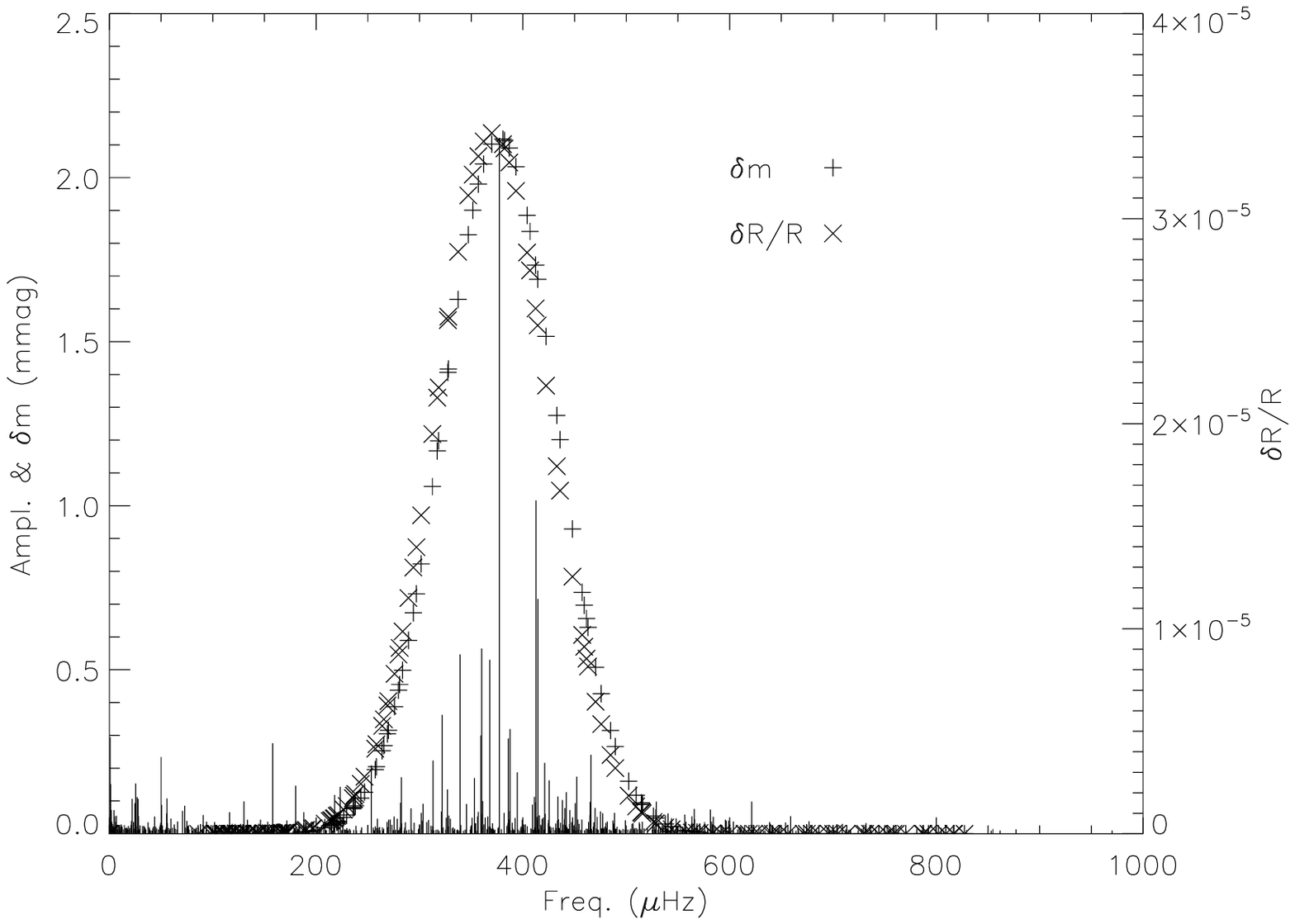}}
 &
 \resizebox{0.48\linewidth}{6.5cm}{\includegraphics{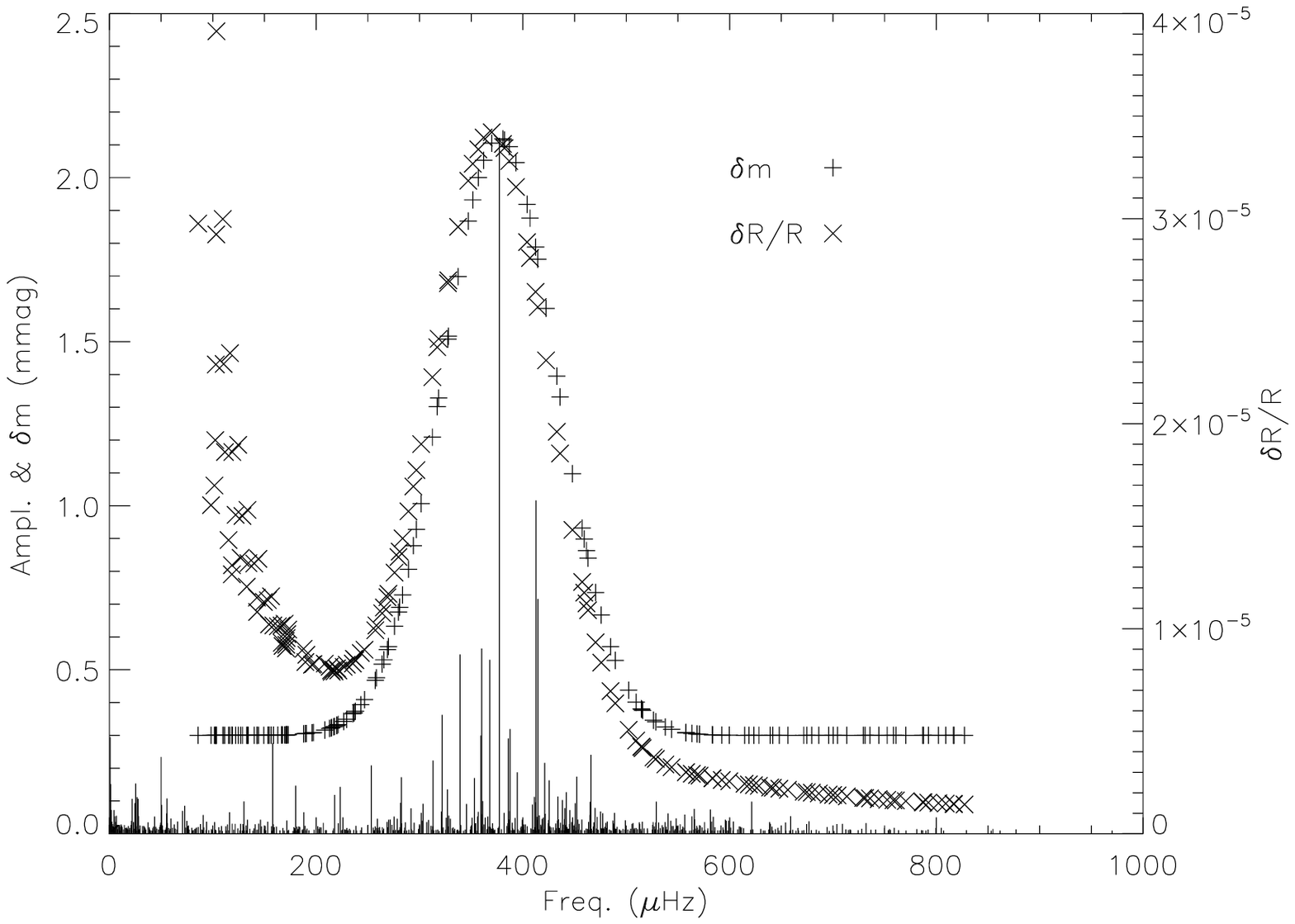}}
 \\
      \end{tabular}
   \caption{Observed amplitude spectrum of HD\,174936 \citep{antonio},
     intrinsic oscillation amplitude ($\delta$m, plus signs;
     scale in the left axis), and relative radius ($\delta R/R$,
       crosses; scale in the right axis).  Left: built with one
     distribution. Right: built with two distributions (see text for
     details).}
   \label{espec}
   \end{figure*}

We use the observed amplitude spectrum for HD\,174936
\citep[Fig.~\ref{espec} and][]{antonio} as an example to illustrate
the procedure here presented. The spectrum shows a maximum oscillation
amplitude of about 2~mmag, which would correspond to a low-degree
mode. Decreasing amplitudes are found at both sides of this highest
amplitude mode. Therefore, we assume that: ``The mode amplitude is
given by a Gaussian distribution centered in the largest amplitude
mode, with the maximum amplitude of this mode, and the variance fixed
by the decrease in amplitude of the modes closest to the highest
one. This intrinsic amplitude is independent of the spherical degree
$\ell$ of the mode''. The oscillation amplitudes of the other modes,
out of the highest amplitude one, would be affected by geometrical
visibility, which is translated in amplitudes lower than the predicted
Gaussian envelope.

In this case, the distribution is centered on 377~$\mu$Hz, with a
maximum amplitude of 2.12~mmag and a square root of the variance
$\sigma=50~\mu$Hz. The assumed oscillation amplitudes are also shown in
Fig.~\ref{espec} (left, plus signs), compared with the observed
amplitude spectrum. We notice that a significant number of observed
amplitudes fall out of the range $[200,500]~\mu$Hz, where our
distribution predicts amplitudes close to zero. Therefore, we built a
modified distribution as the addition of the Gaussian already
explained plus a flat-background at 0.3~mmag for the modes with
negligible amplitude (Fig.~\ref{espec}, right, plus signs). There is
no physical reasons for this election, but only a fitting of the
observations. The value of the background was chosen to take into
account the amplitude of the observed mode at 158.1~$\mu$Hz, and the
distribution is spanned in the range $[100,800]~\mu$Hz well comprising
the canonical range of observed modes for $\delta$ Scuti. We
deliberately leave out modes under 100~$\mu$Hz as they lay far out of
the instability range of $\delta$ Scuti and no model is able to
predict their excitation.

The next step is to relate this intrinsic oscillation amplitude in
luminosity to oscillation amplitude in relative radius. To do so, we
use the definition of the apparent magnitude variation for a certain
wavelength given by \citet{balona}

\begin{equation}
\Delta m_{\lambda} = \Delta S_{\lambda} -2.5 \log e \Delta A/A
\label{eq:mlambda}
\end{equation}

\noindent where $A$ is the projected area of the photosphere, and
$\Delta S_{\lambda}$ is the variation in surface brightness defined as

\begin{equation}
\Delta S_{\lambda} = -2.5 \Delta (\log \bar{F}_{\lambda})
\end{equation}

\noindent with $\bar{F}_{\lambda}$ being the projected flux at a
certain wavelength.

Given that the unknown multiplicative constant is defined by the
variation of the radius, we are only interested in the intrinsic
amplitude of the radial eigenfunction. Therefore, although the
relative variation of the projected surface area $\Delta A/A$ includes
dependencies as the degree $\ell$ of the modes, the viewing angle or
the limb darkening, we may adopt, for our purposes, $\Delta A$ as the
radial variation of the total surface of the star.  In addition, as
the observed CoRoT amplitude spectrum for HD\,174936 was taken in
white light, we will consider $\bar{F}_{\lambda}$ as the total flux at
the surface. Therefore, $\Delta m_{\lambda} = \Delta m = \Delta
M_{bol}$, since the bolometric magnitude ($M_{bol}$) and absolute
magnitude ($m$) differ only in a constant. Including all this in
Eq.~\ref{eq:mlambda} yields

\begin{equation}
\Delta m = -2.5  \log e \left[ \Delta (\ln \sigma T_{eff}^4) - 2{\Delta R \over R}  \right]
\end{equation}

\noindent where $\sigma$ is Stefan Boltzmann constant. Simple calculations give:

\begin{equation}
\Delta m=(-5{\Delta R\over R}-10{\Delta
  T_{eff}\over T_{eff}})\log e
\end{equation}

Therefore, the variations in brightness are a combination of the
variations of the relative radius and the relative flux at the surface
of the star, measured through relative variations of the effective
temperature. The observed variations of the relative radius may
be obtained as

\begin{equation}
{\Delta R\over R}=-{\Delta m\over \log e (5+10dT)}
\end{equation}

\noindent where

\begin{equation}
dT={\delta T_{eff}\over T_{eff}}/{\xi_r\over r}\;\;\; \mathrm{at} \;\;\;r=R
\end{equation}

The negative sign of this equation is related to the definition of the
stellar magnitude and it has no physical origin. We will not take it
into account in the rest of this study, since it has no influence in
the results. Therefore, once the variations in brightness are known,
the relative radius variations can be obtained, as the variable $dT$
is supplied by the non-adiabatic equations. We used the GraCo
non-adiabatic code for this purpose, including the Time Dependent
Convection description \citep[TDC,][]{ahmed05,ma05}, although its
inclusion revealed of no significance in the results. The final
relative radius variations as a function of the frequency of the modes
for the two distributions are displayed also in Fig.~\ref{espec}.

\section{Obtaining the energy balance of the modes for different
  stellar cases}

Once the multiplicative constant $\delta R/R$ scaling all the linear
solutions to their real values is known, the absolute values of the
kinetic energy and the energy required to excite the observed modes
can be estimated. Then, they can be compared to the energy balancing
the star and the energy available in the star, respectively.

To check if the star is able to have thousands of modes excited at
a time, we have to verify 1) if the energy flux is enough to provide
the energy that all these modes need per second, and 2) if the total
kinetic energy accumulated in them does not destroy the equilibrium
structure of the star. We note that quantitative estimations of upper
limits of these energies are enough to achieve the aims of this work.

Three stellar cases have been studied: 1) a representative model
  of the classical Cepheid stellar type, adopting the
largest pulsation amplitude reported (around 2~mag) as a
falsifiability test of our main assumptions, 2) the $\delta$ Scuti
star HD\,174936, and 3) the hybrid $\delta$ Scuti - $\gamma$ Doradus
star HD\,49434.

For the latter two cases, we have used CoRoT and ground-based
observations. In the case of HD\,49434, the publication of the data
analysis is still in preparation and we used here only some general
and preliminary results, which is, however, enough for our purpose.

% here only some general and preliminary
%results, enough for our purposes, are used.

\subsection{A classical Cepheid star}

Our first stellar case applies to the variable stellar type with
the highest amplitude luminosity known to date, the Classical
Cepheid type. This type of stars do not show a large amount of
pulsational modes; however, we are going to use it as a test of
falsifiability of our main hypotheses, since for them, its high
  variability offsets the smaller number of pulsation modes.

We built a representative equilibrium model lying in the
classical Cepheid instability strip \citep{sanbade}. The main physical
characteristics of this model are displayed in Table 1, and are
  similar to one of the thoeretical models studied in
  \cite{bono}. The energy balance of the pulsational modes is then
compared to some absolute values of the star and with some
  theoretical results obtained through non-linear resolutions, as
described in Section~2.

The variation of the relative radius of the fundamental radial
  mode obtained following our procedure ($\Delta R/R=0.16$) is similar
  to the predicted for an equivalent non-linear model
  described in \cite{bono} ($\Delta R/R=0.13$).

% Then we have
%studied the energy balances of the pulsational modes as compared with
%some absolute values of the star (see Section 2).

\subsubsection{Pulsational energy}

The energy balance between gains and losses, for a single mode, in a
complete period of oscillation is given by \citep{unno}

\begin{equation}
W={\pi\over\sigma}\int_0^M{\delta T\over
  T}\delta(\epsilon_N-{1\over\rho}\nabla\cdot\vec F)dM
\end{equation}

This quantity is provided by GraCo and can also be regarded as the
required energy for each mode to be driven during one oscillation
cycle. The scaling of the linear solutions to their real values is
done multiplying Eq. 9 by $A^2_{real}$. As we are interested in an
upper limit for this quantity, we assume that for every mode the
energy balance is positive, i.e. that each mode removes energy from
the star. This is only true for unstable modes, as stable modes
transfer energy to the star. However, in this way, we make sure that
the real amount of energy subtracted by the modes from the star,
regardless of the theory used to obtain it, will be always lower than
our estimated upper limit.

The total energy during a cycle of oscillation is given by the sum of
all the energy interchanged by all the modes in the observed frequency
range. We compared this energy to the luminosity of the star, as a
measure of the available radiation energy per time unit. For Classical
Cepheid stars, we have used the fundamental radial mode and its
overtones. Therefore, the sum of $W \cdot \sigma$ for each mode
($\sigma$ being the frequency of the mode in $Hz$) yields a total
energy per time unit needed by the modes to be overstable of
$10^{35}$~ergs~s$^{-1}$. The luminosity of the star is of the order of
$10^{36}$~ergs~s$^{-1}$, that is, one order of magnitude higher.  As
we expected the real energy value to be even lower than the one
estimated here, we conclude that this extreme case has enough energy
per time unit to unstabilize all observed modes. This result is used
only to legitimize our main assumptions.

%This is the
%expected result, supporting our main assumptions.

\subsubsection{Total kinetic energy}

For the second comparison we have calculated the kinetic energy of
each mode as \citep{unno}

\begin{equation}
E_{kin}={1\over 2}\sigma^2\int_0^M|\vec\xi|^2dM
\end{equation}

Once the kinetic energy of each mode is scaled by its variation of the
relative radius (multiplying by $A^2_{real}$), the total sum of the
kinetic energies of the selected modes gives a total kinetic energy of
the order of $10^{41}$~ergs. \cite{bono} give the same order of
  magnitude for the kinetic energy of the model we are comparing
  with. This value has to be compared also with the total
energy of the star, i.e., the sum of the internal energy of the star,
mass, rotational kinetic energy, turbulent energy, gravitational
energy, etc. The estimation of all these parameters is a complex task,
but the gravitational energy that holds the star together is already
of the order of $10^{49}$~ergs. Therefore, we can infer that the total
kinetic energy of pulsations of the radial modes for the case of
  the Classical Cepheid stars is low enough for the star to bear it
maintaining the hydrostatic equilibrium. Again, our assumptions are
validated with this test.

%Supporting again our assumptions

\subsection{A $\delta$ Scuti star observed by CoRoT (HD\,174936)}

%The first study with unexpected space data results has been done for
%the $\delta$ Scuti star observed by CoRoT HD\,174936.

The $\delta$ Scuti HD\,174936 was the first CoRoT target of this
stellar type (together with HD\,50844) presenting unexpected space
data. The observational spectrum and the relative variation of the
radius have already been presented and explained in Section 2 and
depicted in Fig.~1. The physical characteristics of the model used are
displayed in Table~1.

The total energy per oscillation cycle is given by the sum of
all the energy interchanged by all the modes in the frequency range
$[100,800]~\mu$Hz (a generous gamut for $\delta$ Scutis, as we said
above) and with degrees $\ell=[0,7]$, to be able to account for the
thousands of observed modes. The observation of spherical degrees up
to $\ell=7$ with high-precision photometric time series were predicted
by \cite{yagoda}.

The sum of $W \cdot \sigma$ for each mode yields a total energy
per time unit needed by the modes to be overstable of
$10^{29}$~ergs~s$^{-1}$, for both proposed amplitude distributions
(with and without TDC). The luminosity of the star is of the order of
$10^{34}$~ergs~s$^{-1}$, that is, five orders of magnitude higher.
Consequently, as the real energy value is even lower than the one here
estimated, we conclude that the star has enough energy per time unit
to unstabilize all the observed modes.

On the other hand, the total sum of the kinetic energies of the
selected modes gives a total kinetic energy in the range
$[10^{39},10^{41}]$~ergs, depending on the amplitude distribution. The
gravitational energy holding the star together is of the order of
$10^{48}$~ergs. Therefore, as in the previous case, the total kinetic
energy of pulsations of the thousands of modes is low enough for the
star to bear it maintaining the hydrostatic equilibrium.

\subsection{An hybrid $\gamma$ Doradus - $\delta$ Scuti star observed by CoRoT (HD\,49434)}

The last stellar case studied is an hybrid $\delta$ Scuti - $\gamma$
Doradus star: HD\,49434 which has also been observed by CoRoT. The
analysis of its light-curve is still in progress and the results have
not yet been published. Hence, we use only the preliminary result that
thousands of modes have been detected in the $ \gamma$ Doradus region,
tens of them reported as ``undoubtly'' real (private communication).

Pulsations in the $\delta$ Scuti region of the frequency spectrum have
also been reported \citep{juntenpoyas}. We assume that in this region
the situation is similar to that found for HD\,174936 in order to
study a possible future situation.  \cite{juntenpoyas} did
ground-based multisite observations for this star: we use the highest
amplitude mode reported, to fix the center and maximum amplitude of
our Gaussian distribution. The mode has a photometric amplitude of
2.0~mmag at a frequency of 20~$\mu$Hz. The observed frequencies in the
$\delta$ Scuti region present values around 90~$\mu$Hz. We have
proceed in a way similar to the previous case, but using a double
Gaussian, one for each region presenting pulsational modes. A
flat-background of 0.3~mmag has also been used for a second possible
distribution, as explained in Section~2.

We built a model fulfilling the physical characteristics shown in
\cite{juntenpoyas} (see Table 1.). For this study, frequencies in the
range $[3,900]\mu$Hz, to well cover the $\delta$ Sct and $\gamma$ Dor
range, and degrees in the range $\ell ={0,7}$ have been used.

We obtain that the total energy per time unit needed by the
modes to be overstable is of the order of $10^{30}$~ergs~s$^{-1}$
(only using TDC, more suitable for $\gamma$ Dor targets). The
luminosity of the star is of the order of $10^{34}$~ergs~s$^{-1}$,
that is, four orders of magnitude higher. We conclude again that the
star has enough energy per time unit to unstabilize all the
possible modes. The total sum of the kinetic energies of the modes
selected are in the range $[10^{42},10^{43}]$~ergs, depending on the
amplitude distribution. The gravitational energy that holds the star
together is of the order $10^{49}$~ergs. Again, we conclude that
the total kinetic energy of pulsation of the thousands of modes is
low enough for the star to bear it maintaining its hydrostatic
equilibrium.

\section{Conclusions}

Using a non-adiabatic linear pulsational code, we have related the
observed variation in brightness of Classical Cepheid stars, a
$\delta$ Scuti, and an hybrid $\delta$ Sct-$\gamma$ Dor star
with the intrinsic variation of their relative radius. This has
allowed us to scale theoretical quantities from this linear code of
oscillations to real values of the star, and to make the first
comparison to date, up to our knowledge, of absolute pulsational
quantities with absolute parameters of non-stochastic pulsating stars.

The subsequent analysis of the total kinetic energy of the thousands
of observed modes, and of the required energy to excite them, shows
that all of them have enough energy to sustain their oscillation. Our
study shows that the results yielded by the analyses of the light
curves with the classical tools (standard sine wave fitting) dot not
come into conflict with what it is physically possible.

\acknowledgments

     AM acknowledges financial support from a {\em Juan de la Cierva}
     contract of the Spanish Ministry of Science and Innovation. CRL
     acknowledges an {\em \'Angeles Alvari\~ no} contract under {\em
       Xunta de Galicia}.

\clearpage

%-------------------------------------------------------------

\clearpage

\begin{table}
\begin{center}
\caption{Physical characteristics of the model used}
\label{table1}
\begin{tabular}{cccc}
\tableline\tableline
Type & $M/M_\odot$ & $L/L_\odot$ & $R/R_\odot$ \\
\tableline
Class. Ceph. & 5.00 & 496.5  & 22.49 \\
$\delta$ Sct. &  1.63 & 14.21  & 1.97 \\
Hybrid & 1.60 & 6.80 & 1.58 \\
\tableline
\end{tabular}
\end{center}
\end{table}

\end{document}